\documentclass[aps,prd,twocolumn,superscriptaddress,nofootinbib]{revtex4-1}

\usepackage{latexsym}
\usepackage{amsmath}
\usepackage{amssymb}
\usepackage{amsfonts}
\usepackage{bm}
\usepackage{physics}

\usepackage{color}
\definecolor{purple}{rgb}{0.5,0,0.5}
\definecolor{blue}{rgb}{0.0,0,0.9}
\definecolor{prdblue}{rgb}{0.133,0.118,0.498}
\usepackage[colorlinks=true, pdfstartview=FitV, linkcolor=prdblue, citecolor= prdblue, urlcolor=prdblue]{hyperref}

\usepackage{supertabular} 
\usepackage{placeins}
\usepackage{epsfig}
\usepackage{graphicx}

\begin{document}


\title{Unraveling the nature of the novel $\mathbf{T_{cs}}$ and $\mathbf{T_{c\bar s}}$ tetraquark candidates}


\author{P. G. Ortega}
\email[]{pgortega@usal.es}
\affiliation{Departamento de F\'isica Fundamental, Universidad de Salamanca, E-37008 Salamanca, Spain}
\affiliation{Instituto Universitario de F\'isica 
Fundamental y Matem\'aticas (IUFFyM), Universidad de Salamanca, E-37008 Salamanca, Spain}

\author{D. R. Entem}
\email[]{entem@usal.es}
\affiliation{Departamento de F\'isica Fundamental, Universidad de Salamanca, E-37008 Salamanca, Spain}
\affiliation{Instituto Universitario de F\'isica
Fundamental y Matem\'aticas (IUFFyM), Universidad de Salamanca, E-37008 Salamanca, Spain}

\author{F. Fern\'andez}
\email[]{fdz@usal.es}
\affiliation{Instituto Universitario de F\'isica 
Fundamental y Matem\'aticas (IUFFyM), Universidad de Salamanca, E-37008 Salamanca, Spain}

\author{J. Segovia}
\email[]{jsegovia@upo.es}
\affiliation{Departamento de Sistemas F\'isicos, Qu\'imicos y Naturales, Universidad Pablo de Olavide, E-41013 Sevilla, Spain}


\date{\today}

\begin{abstract}
Using proton-proton collisions at centre-of-mass energies $7$, $8$, and $13$ TeV, with a total integrated luminosity of $9\,\text{fb}^{-1}$, the LHCb collaboration has performed amplitude analyses of the $B^+\to D^+D^-K^+$, $B^+\to D^- D_s^+ \pi^+$ and $B^0\to \bar{D}^0 D_s^+ \pi^-$ decays, observing that new $T_{cs}$ and $T_{c\bar s}$ resonances are required in order to explain the experimental data. These signals could be the first observation of tetraquark candidates that do not contain a heavy quark-antiquark pair; in fact, they consist of four different flavours of quarks, one of which is a doubly charged open-charm state.
We present herein an analysis of the $T_{cs}$ and $T_{c\bar s}$ states, which is an extension of our recently published study of similar $T_{cc}^+$ exotic candidates. Our theoretical framework is a constituent-quark-model-based coupled-channels calculation of $qq^\prime \bar s \bar c$ and $cq\bar s\bar q^{\prime}$ tetraquark sectors for $T_{cs}$ and $T_{c\bar s}$ structures, respectively. We explore the nature, and pole position, of the singularities that appear in the scattering matrix with spin-parity quantum numbers: $J^P=0^\pm$, $1^\mp$, and $2^\pm$. The constituent quark model has been widely used in the heavy quark sector, and thus all model parameters are already constrained from previous works. This makes our predictions robust and parameter-free.
We find many singularities in the solution of various scattering-matrix problems which are either virtual states or resonances, but not bound states. Some of them fit well with the experimental observations of the spin-parity, mass and width of $T_{cs}$ and $T_{c\bar s}$ candidates, and thus tentative assignments are made; however, with caution, because the experimental Breit-Wigner parameters are related to the pole characteristics.
\end{abstract}


\maketitle


\section{INTRODUCTION}
\label{sec:introduction}

One century of fundamental research in atomic and nuclear physics has shown that all matter is corpuscular, with the atoms that comprise us, themselves containing a dense nuclear core, which is composed of protons and neutrons, collectively called nucleons, which are members of a broader class of femtometre-scale particles, named hadrons. In working towards an understanding of hadrons, we have discovered that they are complicated bound-states of quarks and gluons whose interactions are described by a Poincar\'e invariant quantum non-Abelian gauge field theory; namely, Quantum Chromodynamics (QCD).

QCD has three salient features: (i) colour confinement, which means that isolated quarks and gluons have never been observed, preferring instead to be grouped into colourless hadrons; (ii) asymptotic freedom characterizing the strength of the interaction between quarks and gluons, which increases as the energy scale decreases, making perturbative calculations in terms of the coupling constant invalid in the low energy region where hadrons coexist; and (iii) chiral symmetry and its spontaneous breaking, which basically mean that the QCD Lagrangian exhibits exact $SU(3)_L \otimes SU(3)_R$ chiral symmetry when the current quark masses of the up, down and strange quarks vanish, however, it is spontaneously broken to the $SU(3)_V$. In addition, other (approximate) symmetries can be used to characterise the dynamics of QCD. For example, at small current quark masses, QCD exhibits the approximate $SU(3)$ flavor symmetry. On the other hand, when the masses of the charm, bottom and even top quarks are very large, QCD shows non-relativistic dynamics and approximate heavy quark spin symmetry.

The information about QCD dynamics in the non-perturbative regime is encoded in the observable properties of hadrons. This complexity makes hadron spectroscopy, the collection of readily accessible states constituted from gluons and quarks, the starting point for all further investigations. A very successful classification scheme for hadrons in terms of their valence quarks and antiquarks was proposed independently by Murray Gell-Mann~\cite{Gell-Mann:1964ewy} and George Zweig~\cite{Zweig:1964CERN} in 1964. This classification, called the quark model, basically divides hadrons into two large families: mesons and baryons. These are quark-antiquark and three-quark bound-states, respectively.\footnote{For further details, the interested reader is referred to the Particle Data Group and its topical mini-review on the subject~\cite{ParticleDataGroup:2022pth}.} However, QCD allows for the existence of more complex structures, generically called exotic hadrons or simply exotics. These include tetra-, penta- and even hexa-quark systems, hadronic states with active gluonic degrees of freedom (hybrids), and even bound states consisting only of gluons (glueballs).

Exotic hadrons have been systematically searched for since the 1960s in numerous experiments around the globe, without success until the beginning of this millennium, when many new hadrons that do not exhibit the expected properties of ordinary mesons and baryons were discovered. These hadrons belong mostly to the heavy quark sector and are collectively known as \emph{XYZ} states. An enormous effort devoted to unravel the nature of these exotic hadrons has been deployed using a wide variety of theoretical approaches. In fact, one can already find many comprehensive reviews on the subject in the literature~\cite{Dong:2020hxe, Chen:2016spr, Guo:2017jvc, Liu:2019zoy, Yang:2020atz, Dong:2021bvy, Chen:2021erj, Mai:2022eur, Meng:2022ozq, Chen:2022asf, Guo:2022kdi, Ortega:2020tng}.

The two most promising theoretical approaches to the \emph{XYZ} states are effective field theories (EFTs) and lattice-regularised QCD (Lattice-QCD). EFTs derived directly from QCD, such as non-relativistic QCD (NRQCD)~\cite{Caswell:1985ui, Bodwin:1994jh} or potential non-relativistic QCD (pNRQCD)~\cite{Pineda:1997bj, Brambilla:1999xf} (for some reviews see Refs.~\cite{Brambilla:2004jw, Pineda:2011dg}), efficiently disentangle the dynamics of the heavy quarks from that of the light degrees of freedom in a model-independent way. On the other hand, the fully relativistic dynamics can be treated without approximations using lattice gauge theories (see, e.g., Ref.~\cite{Dudek:2007wv} for a standard lattice heavy quarkonium spectrum). Heavy quark calculations within these two approaches have experienced a considerable progress for states away from the lowest open-flavor meson-meson decay threshold. However, most of the \emph{XYZ} states appear in the threshold region and then remain problematic for both EFTs and Lattice-QCD~\cite{Brambilla:2010cs, Gray:2005ur, Meinel:2009rd, Burch:2009az, Meinel:2010pv, Dowdall:2011wh, Lewis:2012ir, Wurtz:2015mqa}. Taken together, this explains why many of our theoretical expectations in exotic heavy hadrons still rely on phenomenological potential models. Potential formulations have successfully described the heavy quark-antiquark system since the early days of charmonium and bottomonium (see {\it e.g.}~\cite{Eichten:1978tg, Eichten:1979ms, Gupta:1982kp, Gupta:1983we, Gupta:1984jb, Gupta:1984um, Kwong:1987ak, Kwong:1988ae, Barnes:1996ff, Ebert:2002pp, Radford:2007vd, Eichten:2007qx, Danilkin:2009hr, Ferretti:2013vua, Godfrey:2015dia}). Moreover, the predictions within this formalism of heavy quarkonium properties related to decays and reactions have proved very valuable for experimental searches. Finally, the simplicity of extending the quark model to describe exotic systems makes this framework a very suitable one for exploratory purposes.

Therefore, considering all of the above, in this manuscript we analyse the nature of the recently discovered $T_{c\bar s}$ and $T_{cs}$ states using a constituent quark model (CQM)~\cite{Vijande:2004he, Segovia:2013wma} which is widely used in the heavy quark sector, by studying their spectra~\cite{Segovia:2008zz, Segovia:2015dia, Segovia:2016xqb, Yang:2019lsg, Ortega:2020uvc}, their electromagnetic, weak and strong decays and reactions~\cite{Segovia:2009zz, Segovia:2011zza, Segovia:2012yh, Segovia:2013kg, Segovia:2014mca}, their possible compact multiquark components~\cite{Yang:2018oqd, Yang:2020fou, Yang:2020twg, Yang:2021zhe, Yang:2023mov} and also their potential coupling to meson-meson thresholds~\cite{Ortega:2009hj, Ortega:2016pgg, Ortega:2018cnm, Ortega:2021xst, Ortega:2021fem}. The advantage of using an approach with a relatively long history is that all model parameters are already constrained by previous works. Consequently, from this perspective, we present a parameter-free calculation of the $T_{c\bar s}$ and $T_{cs}$ states, which is also an extension of our recently published analysis of similar $T_{cc}^+$ exotic candidates~\cite{Ortega:2022efc}.

The LHCb collaboration, using proton-proton collision data taken at centre-of-mass energies of $7$, $8$, and $13$ TeV, with an integrated luminosity of $9\,\text{fb}^{-1}$, has recently performed amplitude analyses of the decays $B^+\to D^+D^-K^+$~\cite{LHCb:2020pxc, LHCb:2020bls}, $B^+\to D^- D_s^+ \pi^+$ and $B^0\to \bar{D}^0 D_s^+ \pi^-$~\cite{LHCb:2022xob, LHCb:2022bkt}. For the $B^+\to D^+D^-K^+$ reaction, in order to obtain good agreement with the experimental data, it turns out to be necessary to include new spin-$0$ and spin-$1$ $T_{cs}$ resonances in the $D^-K^+$ channel. These signals may constitute the first observation of exotic hadrons not containing a heavy quark-antiquark pair and, moreover, they could be the first experimental detection of tetraquark candidates with four different flavours of quarks $ud\bar{s}\bar{c}$. On the other hand, the enhancement in the $D_s^+\pi^+$ invariant mass of the $B^+\to D^- D_s^+ \pi^+$ decay may indicate the first observation of a doubly charged open-charm tetraquark state with minimal quark content $c\bar s u\bar d$; whereas the one observed in the $D_s^+\pi^-$ channel of the $B^0\to \bar{D}^0 D_s^+ \pi^-$ reaction is interpreted as the $T_{c\bar s}$ neutral partner of an isospin triplet. Both $T_{c\bar s}$ candidates are found to have isospin $1$ and spin-parity $J^P=0^+$.

The above announcements made by the LHCb collaboration triggered a lot of theoretical work using a wide variety of approaches. Regarding the $T_{cs}$ candidates, one can mention interpretations of these states as either compact tetraquarks or hadron molecules using QCD sum rules~\cite{Chen:2020aos, Agaev:2020nrc, Zhang:2020oze, Albuquerque:2020ugi, Wang:2020xyc}, nonrelativistic and (extended) relativized quark models with different types of quark--(anti-)quark interactions~\cite{Cheng:2020nho, Liu:2020nil, He:2020btl, Xue:2020vtq, He:2020jna, Wang:2020prk, Garcilazo:2020bgc, Karliner:2020vsi, Yang:2021izl, Tan:2020cpu, Hu:2020mxp}, and effective field theories~\cite{Molina:2010tx, Molina:2020hde, Dong:2020rgs}, but they can be also interpreted as triangle singularities~\cite{Liu:2020orv, Burns:2020epm}. In addition, their decay and production properties have been studied in the literature~\cite{Xiao:2020ltm, Chen:2020eyu, Burns:2020xne}. With respect to $T_{c\bar s}$ signals, the literature is more limited; to mention only a few, the reader is referred to Refs.~\cite{Liu:2022hbk, Wei:2022wtr, Chen:2022svh, Yue:2022mnf, Molina:2022jcd, Yang:2023evp, Dmitrasinovic:2023eei, Lian:2023cgs}.

The manuscript is structured as follows. After this introduction, the theoretical framework is briefly presented in section~\ref{sec:theory}. Section~\ref{sec:results} is mainly devoted to the analysis and discussion of our theoretical results. Finally, we summarise and draw some conclusions in Sec.~\ref{sec:summary}.


\section{THEORETICAL FORMALISM}
\label{sec:theory} 

\subsection{Naive quark model}

Constituent light quark masses and Goldstone-boson exchanges, which are consequences of dynamical chiral symmetry breaking in QCD, together with the perturbative one-gluon exchange (OGE) and a non-perturbative confinement interaction, are the main parts of our constituent quark model~\cite{Vijande:2004he, Segovia:2013wma}.

A simple Lagrangian invariant under chiral transformations can be written in the following form~\cite{Diakonov:2002fq}:
\begin{equation}
{\mathcal L} = \bar{\psi}(i\, {\slash\!\!\! \partial} - M(q^{2}) U^{\gamma_{5}}) \,\psi \,,
\end{equation}
where $M(q^2)$ is the dynamical (constituent) quark mass and $U^{\gamma_5} = e^{i\lambda _{a}\phi ^{a}\gamma _{5}/f_{\pi}}$ is the matrix of Goldstone-boson fields that can be expanded as
\begin{equation}
U^{\gamma _{5}} = 1 + \frac{i}{f_{\pi}} \gamma^{5} \lambda^{a} \pi^{a} - \frac{1}{2f_{\pi}^{2}} \pi^{a} \pi^{a} + \ldots
\end{equation}
The first term of the expansion generates the constituent quark mass while the second term gives rise to a one-boson exchange interaction between the quarks. The main contribution of the third term comes from the two-pion exchange interaction, simulated by means of a scalar-meson exchange potential.

It is well known that multi-gluon exchanges produce an attractive linearly rising potential proportional to the distance between infinite-heavy quarks~\cite{Bali:2000gf}. However, sea quarks are also important ingredients of the strong interaction dynamics that contribute to the screening of the rising potential at low momenta and eventually to the breaking of the quark-antiquark binding string~\cite{Bali:2005fu}. Our constituent quark model attempts to mimic this behavior by implementing the following expression:
\begin{equation}
V_{\rm CON}(\vec{r}\,)=\left[-a_{c}(1-e^{-\mu_{c}r})+\Delta \right] (\vec{\lambda}_{q}^{c}\cdot\vec{\lambda}_{\bar{q}}^{c}) \,,
\label{eq:conf}
\end{equation}
where $a_{c}$ and $\mu_{c}$ are model parameters. At short distances, this potential presents a linear behavior with an effective confinement strength, $\sigma=-a_{c}\,\mu_{c}\,(\vec{\lambda}^{c}_{i}\cdot \vec{\lambda}^{c}_{j})$, while it becomes constant at large distances, with a threshold defined by
\begin{equation}
V_{\rm thr} = \{-a_{c}+\Delta\}(\vec{\lambda}^{c}_{i}\cdot \vec{\lambda}^{c}_{j}).
\end{equation}

Finally, the dynamics of the bound state system is expected to be governed by QCD perturbative effects at short inter-quark distances. This is accounted for by the one-gluon exchange potential derived from the following vertex Lagrangian
\begin{equation}
{\mathcal L}_{qqg} = i\sqrt{4\pi\alpha_{s}} \, \bar{\psi} \gamma_{\mu} 
G^{\mu}_c \lambda^c \psi \,,
\label{eq:Lqqg}
\end{equation}
with $\alpha_{s}$ an effective scale-dependent strong coupling constant which allows us to consistently describe light, strange and heavy quark sectors, and whose explicit expression can be found in, \emph{e.g.}, Ref.~\cite{Segovia:2008zz}.

Detailed expressions for all the potentials and the values of the model parameters can be found in Ref.~\cite{Vijande:2004he}, updated in Ref.~\cite{Segovia:2008zz}. We would like to point out here that the interaction terms between light-light, light-heavy and heavy-heavy quarks are not the same in our formalism, \emph{i.e.} while Goldstone-boson exchanges are considered when the two quarks are light, they do not appear in the other two configurations: light-heavy and heavy-heavy; however, the one-gluon exchange and confinement potentials are blinded in flavor and so they affect all the cases. Finally, the meson eigenenergies and eigenstates are obtained by solving the Schr\"odinger equation using the Gaussian Expansion Method~\cite{Hiyama:2003cu} which provides sufficient accuracy and simplifies the subsequent evaluation of the required matrix elements.

\subsection{Resonating group method}

The aforementioned CQM specifies the microscopic interaction between the constituent quarks and antiquarks. To describe the interaction at the meson level, we use the Resonating Group Method~\cite{Wheeler:1937zza}, where mesons are considered as quark-antiquark clusters and an effective cluster-cluster interaction emerges from the underlying quark(antiquark) dynamics.

We assume that the wave function of a system composed of two mesons $A$ and $B$ can be written as\footnote{Note that, for the simplicity of the discussion presented here, we have omitted the spin-isospin wave function, the product of the two colour singlets and the wave function describing the centre-of-mass motion.}
\begin{equation}
\langle \vec{p}_{A} \vec{p}_{B} \vec{P} \vec{P}_{\rm c.m.} | \psi 
\rangle = {\cal A}\left[\phi_{A}(\vec{p}_{A}) \phi_{B}(\vec{p}_{B}) 
\chi_{\alpha}(\vec{P})\right] \,,
\label{eq:wf}
\end{equation}
where ${\cal A}$ is the full antisymmetric operator, $\phi_{C}(\vec{p}_{C})$ is the wave function of a general meson $C$ calculated in the naive quark model, and $\vec{p}_{C}$ is the relative momentum between the quark and antiquark of the meson $C$. The wave function taking into account the relative motion of the two mesons is $\chi_\alpha(\vec{P})$, where $\alpha$ denotes the set of quantum numbers needed to uniquely define a particular partial wave.

The projected Schr\"odinger equation for the relative wave function can be written as follows:
\begin{align}
&
\left(\frac{\vec{P}^{\prime 2}}{2\mu}-E \right) \chi_\alpha(\vec{P}') + \sum_{\alpha'}\int \Bigg[ {}^{\rm RGM}V_{D}^{\alpha\alpha'}(\vec{P}',\vec{P}_{i}) + \nonumber \\
&
+ {}^{\rm RGM}V_{R}^{\alpha\alpha'}(\vec{P}',\vec{P}_{i}) \Bigg] \chi_{\alpha'}(\vec{P}_{i})\, d\vec{P}_{i} = 0 \,,
\label{eq:Schrodinger}
\end{align}
where $E$ is the energy of the system. The direct potential ${}^{\rm RGM}V_{D}^{\alpha\alpha '}(\vec{P}',\vec{P}_{i})$ can be written as
\begin{align}
&
{}^{\rm RGM}V_{D}^{\alpha\alpha '}(\vec{P}',\vec{P}_{i}) = \sum_{i\in A, j\in B} \int d\vec{p}_{A'} d\vec{p}_{B'} d\vec{p}_{A} d\vec{p}_{B} \times \nonumber \\
&
\times \phi_{A'}^{\ast}(\vec{p}_{A'}) \phi_{B'}^{\ast}(\vec{p}_{B'})
V_{ij}^{\alpha\alpha '}(\vec{P}',\vec{P}_{i}) \phi_{A}(\vec{p}_{A}) \phi_{B}(\vec{p}_{B})  \,.
\end{align}
where $V_{ij}^{\alpha\alpha '}$ is the CQM potential between the quark $i$ and the quark $j$ of the mesons $A$ and $B$, respectively.

The quark rearrangement potential ${}^{\rm RGM}V_{R}^{\alpha\alpha'}(\vec{P}',\vec{P}_{i})$ represents a natural way to connect meson-meson channels with the same quark content, and it is given by
\begin{align}
&
{}^{\rm RGM}V_{R}^{\alpha\alpha'}(\vec{P}',\vec{P}_{i}) = \sum_{i\in A, j\in B}\int d\vec{p}_{A'}
d\vec{p}_{B'} d\vec{p}_{A} d\vec{p}_{B} d\vec{P} \phi_{A'}^{\ast}(\vec{p}_{A'}) \times \nonumber \\
&
\times  \phi_{B'}^{\ast}(\vec{p}_{B'})
V_{ij}^{\alpha\alpha '}(\vec{P}',\vec{P}) P_{mn} \left[\phi_{A}(\vec{p}_{A}) \phi_{B}(\vec{p}_{B}) \delta^{(3)}(\vec{P}-\vec{P}_{i}) \right] \,,
\label{eq:Kernel}
\end{align}
where $P_{mn}$ is the operator that exchanges the quark $m$ of $A$ with the quark $n$ of $B$.

From Eq.~\eqref{eq:Schrodinger}, we derive a set of coupled Lippmann-Schwinger equations of the form
\begin{align}
T_{\alpha}^{\alpha'}(E;p',p) &= V_{\alpha}^{\alpha'}(p',p) + \sum_{\alpha''} \int
dp''\, p^{\prime\prime2}\, V_{\alpha''}^{\alpha'}(p',p'') \nonumber \\
&
\times \frac{1}{E-{\cal E}_{\alpha''}(p^{''})}\, T_{\alpha}^{\alpha''}(E;p'',p) \,,
\end{align}
where $V_{\alpha}^{\alpha'}(p',p)$ is the projected potential containing the direct and rearrangement potentials, and ${\cal E}_{\alpha''}(p'')$ is the energy corresponding to a momentum $p''$, written in the non-relativistic case as	
\begin{equation}
{\cal E}_{\alpha}(p) = \frac{p^2}{2\mu_{\alpha}} + \Delta M_{\alpha} \,.
\end{equation}
Here, $\mu_{\alpha}$ is the reduced mass of the $(AB)$-system corresponding to the channel $\alpha$, and $\Delta M_{\alpha}$ is the difference between the threshold of the $(AB)$-system and the one we use as a reference.

We solve the coupled Lippmann-Schwinger equations using the matrix-inversion method proposed in Ref.~\cite{Machleidt:1003bo}, but generalised to include channels with different thresholds. Once the $T$-matrix is computed, we determine the on-shell part which is directly related to the scattering matrix. In the case of non-relativistic kinematics, it can be written as
\begin{equation}
S_{\alpha}^{\alpha'} = 1 - 2\pi i 
\sqrt{\mu_{\alpha}\mu_{\alpha'}k_{\alpha}k_{\alpha'}} \, 
T_{\alpha}^{\alpha'}(E+i0^{+};k_{\alpha'},k_{\alpha}) \,,
\end{equation}
where $k_{\alpha}$ is the on-shell momentum for channel $\alpha$.

Our aim is to explore the existence of states above and below thresholds within the same formalism. Thus, we have to continue analytically all the potentials and kernels for complex momenta in order to find the poles of the $T$-matrix in any possible Riemann sheet.


\section{RESULTS}
\label{sec:results}

A detailed discussion of the peculiarities of our coupled-channels calculation will be given in the following subsections. However, a comment on the theoretical uncertainty is in order here. There are two types of theoretical uncertainties in our results: one is intrinsic to the numerical algorithm and the other is related to the way the model parameters are fixed. The numerical error is negligible and, as mentioned above, the model parameters are adjusted to reproduce a certain number of hadron observables within a determinate range of agreement with experiment. It is therefore difficult to assign an error to these parameters and consequently to the quantities calculated using them.
In order to analyze the uncertainty of the calculations presented in this manuscript, we will estimate the error of the pole properties by varying the strength of our potentials by $\pm10\%$.

\subsection{Nature of $T_{cs}$ states}

The LHCb collaboration carried out an amplitude analysis of $B^+\rightarrow D^+D^-K^+$ decays at the end of 2020 using proton-proton collision data taken at $\sqrt{s}=7$, $8$, and $13$ TeV, with an integrated luminosity of $9\,\text{fb}^{-1}$~\cite{LHCb:2020pxc, LHCb:2020bls}. New spin-$0$ and spin-$1$ charm-strange resonances in the $D^-K^+$ channel had to be included to obtain good agreement with the experimental data. Their Breit-Wigner parameters are
\begin{align}
T_{cs0}(2900)^0: \hspace*{0.2cm} M &= (2866 \pm 7 \pm 2) \,\text{MeV/c}^2 \,, \nonumber \\ \Gamma &= (57 \pm 12 \pm 4) \,\text{MeV} \,, \label{eq:Tcs0} \\[1ex] \nonumber
T_{cs1}(2900)^0: \hspace*{0.2cm} M &= (2904 \pm 5 \pm 1) \,\text{MeV/c}^2 \,, \nonumber \\ \Gamma &= (110 \pm 11 \pm 4) \,\text{MeV} \label{eq:Tcs1} \,,
\end{align}
where the first uncertainties are statistical and the second systematic. With an overwhelming significance, these resonances constitute the first observation of exotic hadrons with open flavor, not containing a heavy quark-antiquark pair and also of a possible tetraquark candidate with four different flavors of quarks: $ud\bar{s}\bar{c}$.

We perform a coupled-channels calculation in charged basis\footnote{The charged basis is selected in this case because there is no experimental evidence of the isospin content of the $T_{cs}$ states.} of the $J^P=0^+$, $1^-$ and $2^+$ $q q'\bar s \bar c$ sectors in which the $D^-K^+$ discovery-channel can be measured. We include the following meson-meson thresholds in the calculation:\footnote{In parenthesis the mass of the threshold in MeV/c$^2$.} $\bar D^0K^{0}$ (2362.45), $D^-K^+$ (2363.34), $D^{*-}K^{*+}$ (2901.92) and $\bar D^{*0}K^{*0}$ (2902.40). Besides the direct interaction between $\bar D^{(*)}K^{(*)}$ pairs, we have to consider exchange diagrams to deal with indistinguishable quarks from different mesons in the molecule. Moreover, the decay width of the strange vector meson is large enough to be included in the calculation. The experimental values of the widths reported in Ref.~\cite{ParticleDataGroup:2022pth} for the neutral and charged partners, respectively, are $\Gamma_{K^{\ast0}}=47.3\,\text{MeV}$ and $\Gamma_{K^{\ast\pm}}=51.4\,\text{MeV}$.

With all the above, our calculation yields the information shown in Table~\ref{tab:Tcs1}. The first observation is that many poles appear in the scattering matrix; they are either virtual or resonance states but we do not find bound states. With all due caution, as we should not compare Breit-Wigner resonance parameters with pole positions, a tentative assignment of the $T_{cs0}(2900)^0$ experimental signal would be the second state with quantum numbers $J^P=0^+$. Its mass and width are $2902\,\text{MeV/c}^2$ and $51\,\text{MeV}$, respectively; both compare well with the corresponding experimental values shown in Eq.~\eqref{eq:Tcs0}. It is worth noting that a very similar state appears in the $J^P=2^+$ channel; however, there is a resonance close (at $2.922$ GeV/c$^2$) to it which has a large decay width that could interfere with the experimental signal. We find a possible candidate of the $T_{cs1}(2900)^0$ signal in the $J^P=1^-$ channel with pole parameters $2888\,\text{MeV/c}^2$ and $190\,\text{MeV}$, and whose nature seems to be virtual. Notice that the $1^-$ $\bar D^*K^*$ molecule is in a relative $P$-wave, whereas it is in a relative $S$-wave for the $0^+$ and $2^+$ sectors. The theoretical width is twice the experimental value, see Eq.~\eqref{eq:Tcs1}, but this could be due to the fact that the experimentalists simply perform cross section fits and do not derive the pole structure that produces the signals.

\begin{turnpage}
\squeezetable
\begin{table}[!t]
\caption{\label{tab:Tcs1} Coupled-channels calculation of the $J^P=0^+$, $1^-$ and $2^+$ $q q'\bar s \bar c$ sectors ($T_{cs}$ states) in which the $D^-K^+$ discovery-channel can be measured. We include the following meson-meson thresholds in the calculation (in parenthesis the threshold's mass in MeV/c$^2$): $\bar D^0K^{0}$ (2362.45), $D^-K^{+}$ (2363.34), $D^{*-}K^{*+}$ (2901.92) and $\bar D^{*0}K^{*0}$ (2902.40). Errors are estimated by varying the strength of the potential by $\pm10\%$. \emph{$1^{st}$ column:} Pole's quantum numbers; \emph{$2^{nd}$ column:} Pole's mass in MeV/c$^2$; \emph{$3^{rd}$ column:} Pole's width in MeV; \emph{$4^{th}$ column:} Refers to $\bar D^0K^{0}$, $D^-K^{+}$, $D^{*-}K^{*+}$ and $\bar D^{*0}K^{*0}$ Riemann sheets, respectively, with $F$ meaning first and $S$ second; \emph{$5^{th}-6^{th}$ columns:} Isospin channel probabilities in \%; \emph{$7^{th}-10^{th}$ columns:} Channel probabilities in \%; \emph{$11^{th}-14^{th}$ columns:} Branching ratios in \%.}
\begin{ruledtabular}
\begin{tabular}{lllcllllllllll}
$J^P$ & $M_{\text{pole}}$ & $\Gamma_{\text{pole}}$ & RS & ${\cal P}_{I=0}$ & ${\cal P}_{I=1}$  & ${\cal P}_{\bar D^0K^0}$ & ${\cal P}_{D^-K^+}$ & ${\cal P}_{D^{*-}K^{*+}}$ & ${\cal P}_{\bar D^{*0}K^{*0}}$ & ${\cal B}_{\bar D^0K^0}$ & ${\cal B}_{D^-K^+}$ & ${\cal B}_{D^{*-}K^{*+}}$ & ${\cal B}_{\bar D^{*0}K^{*0}}$ \\[1ex]
\hline
$ 0^+$ & $2340.0 \pm 0.6$ & $28.8_{-0.7}^{+0.5}$  & (S,S,F,F)  & $96_{-3}^{+2}$  & $4_{-2}^{+3}$   & $57_{-4}^{+5}$  & $33_{-6}^{+5}$  & $5.4 \pm 0.6$  & $4.3_{-0.3}^{+0.2}$   & $0 \pm 0$  & $0 \pm 0$  & $0 \pm 0$  & $0 \pm 0$  \\
       & $2901.9_{-0.8}^{+0.5}$ & $51_{-1}^{+0}$  & (S,S,S,S)  & $22_{-13}^{+7}$  & $78_{-7}^{+13}$   & $0.5_{-0}^{+0.2}$  & $0.1_{-0}^{+0.1}$  & $54_{-34}^{+1}$  & $45_{-1}^{+34}$   & $72_{-12}^{+2}$  & $28_{-14}^{+12}$  & $0_{-0}^{+7}$  & $0_{-0}^{+4}$  \\[1ex]
$ 1^-$ & $2887.7_{-0.4}^{+0.3}$ & $189.5_{-0.6}^{+0.4}$  & (S,S,S,S)  & $10_{-3}^{+4}$  & $90_{-4}^{+3}$   & $0.7_{-0.4}^{+0.6}$  & $9.2 \pm 0.8$  & $51.0_{-0.5}^{+0.9}$  & $39.1_{-0.7}^{+0.1}$   & $12 \pm 5$  & $88 \pm 5$  & $0 \pm 0$  & $0 \pm 0$  \\
       & $3010 \pm 2$ & $257_{-20}^{+23}$  & (S,S,S,S)  & $100\pm0$  & $0\pm0$   & $2.44 \pm 0.06$  & $2.51 \pm 0.06$  & $48.5_{-0.1}^{+0}$  & $46.6_{-0}^{+0.1}$   & $6.0_{-0.4}^{+0.5}$  & $6.2_{-0.5}^{+0.4}$  & $44.1_{-0.6}^{+0.5}$  & $43.7_{-0.4}^{+0.5}$  \\[1ex]
$ 2^+$ & $2902_{-2}^{+1}$ & $49.4_{-0.3}^{+0.2}$  & (S,S,S,S)  & $14_{-12}^{+30}$  & $86_{-30}^{+12}$   & $0.05_{-0.01}^{+0.03}$  & $0.04_{-0.02}^{+0.03}$  & $56_{-3}^{+5}$  & $44_{-5}^{+3}$   & $48_{-46}^{+2}$  & $50_{-48}^{+2}$  & $0_{-0}^{+55}$  & $0_{-0}^{+40}$  \\
       & $2922 \pm 2$ & $287_{-7}^{+8}$  & (S,S,S,S)  & $99.98_{-0.01}^{+0}$  & $0.02_{-0}^{+0.01}$   & $5.26_{-0.01}^{+0.02}$  & $5.28_{-0.01}^{+0.02}$  & $45.45_{-0.03}^{+0.04}$  & $44.01_{-0.08}^{+0.05}$   & $11.8 \pm 0.5$  & $11.9_{-0.4}^{+0.6}$  & $38.6 \pm 0.5$  & $37.6_{-0.5}^{+0.6}$  \\
       & $3160 \pm 1$ & $522_{-33}^{+39}$  & (S,S,S,S)  & $100 \pm 0$  & $0 \pm 0$   & $2.1_{-0.1}^{+0.3}$  & $2.1_{-0.1}^{+0.3}$  & $48.4_{-0.3}^{+0.1}$  & $47.4_{-0.3}^{+0.1}$   & $3.4 \pm 0.2$  & $3.5 \pm 0.2$  & $46.7_{-0.3}^{+0.2}$  & $46.5_{-0.3}^{+0.2}$  \\
\end{tabular}
\end{ruledtabular}

\caption{\label{tab:Tcs2} Coupled-channels calculation of the of the $J^P=0^-$, $1^+$ and $2^-$ $q q'\bar s \bar c$ sectors ($T_{cs}$ states) including the following meson-meson thresholds in the calculation (in parenthesis the threshold's mass in MeV/c$^2$): $D^{*-}K^{+}$ (2503.94), $\bar D^{*0}K^{0}$ (2504.46), $\bar D^{0}K^{*0}$ (2760.39), $D^{-}K^{*+}$ (2761.32), $D^{*-}K^{*+}$ (2901.92) and $\bar D^{*0}K^{*0}$ (2902.40). Errors are estimated by varying the strength of the potential by $\pm10\%$. 
\emph{$1^{st}$ column:} Pole's quantum numbers; \emph{$2^{nd}$ column:} Pole's mass in MeV/c$^2$; \emph{$3^{rd}$ column:} Pole's width in MeV; \emph{$4^{th}$ column:} Refers to $D^{*-}K^{+}$, $\bar D^{*0}K^{0}$, $\bar D^{0}K^{*0}$, $D^{-}K^{*+}$, $D^{*-}K^{*+}$ and $\bar D^{*0}K^{*0}$ Riemann sheets, respectively, with $F$ meaning first and $S$ second; \emph{$5^{th}-6^{th}$ columns:} Isospin channel probabilities in \%; \emph{$7^{th}-12^{th}$ columns:} Channel probabilities in \%; \emph{$13^{th}-18^{th}$ columns:} Branching ratios in \%.
}
\squeezetable
\begin{ruledtabular}
\begin{tabular}{lllcllllllllllllll}
$J^P$ & $M_{\text{pole}}$ & $\Gamma_{\text{pole}}$ & RS & ${\cal P}_{I=0}$ & ${\cal P}_{I=1}$  & ${\cal P}_{D^{*-}K^+}$ & ${\cal P}_{\bar D^{*0}K^0}$ & ${\cal P}_{\bar D^0K^{*0}}$ & ${\cal P}_{D^-K^{*+}}$ & ${\cal P}_{D^{*-}K^{*+}}$ & ${\cal P}_{\bar D^{*0}K^{*0}}$ & ${\cal B}_{D^{*-}K^+}$ & ${\cal B}_{\bar D^{*0}K^0}$ & ${\cal B}_{\bar D^0K^{*0}}$ & ${\cal B}_{D^-K^{*+}}$ & ${\cal B}_{D^{*-}K^{*+}}$ & ${\cal B}_{\bar D^{*0}K^{*0}}$ \\[1ex]
\hline
$ 0^-$ & $2744.6_{-0.2}^{+0.1}$ & $77.45 \pm 0.03$  & (S,S,S,S,F,F)  & $74.7 \pm 0.7$  & $25.3 \pm 0.7$   & $37.8_{-0.7}^{+0.6}$  & $8.3 \pm 0.3$  & $19.6 \pm 0.3$  & $1.5_{-0.1}^{+0.2}$  & $26.7 \pm 0.4$  & $5.97_{-0.08}^{+0.07}$   & $90.0 \pm 0.8$  & $10.0 \pm 0.8$  & $0 \pm 0$  & $0 \pm 0$  & $0 \pm 0$  & $0 \pm 0$  \\
       & $3018_{-11}^{+7}$ & $378 \pm 3$  & (S,S,S,S,S,S)  & $99.97_{-0.03}^{+0.02}$  & $0.03_{-0.01}^{+0.03}$   & $15_{-3}^{+1}$  & $15_{-2}^{+1}$  & $10_{-1}^{+0}$  & $10_{-0}^{+1}$  & $26 \pm 2$  & $25_{-3}^{+2}$   & $18_{-3}^{+2}$  & $17_{-2}^{+3}$  & $13.1_{-0.3}^{+0.2}$  & $13.0_{-0.6}^{+0.7}$  & $20_{-2}^{+1}$  & $19 \pm 2$  \\[1ex]
$ 1^+$ & $2488 \pm 3$ & $0.04\pm0.01$  & (S,S,F,F,F,F)  & $1_{-0}^{+3}$  & $99_{-3}^{+0}$   & $57_{-3}^{+10}$  & $42_{-11}^{+3}$  & $0.3_{-0.1}^{+0.4}$  & $0.13_{-0.07}^{+0.01}$  & $0.3_{-0.1}^{+0}$  & $0.6_{-0.2}^{+0.4}$   & $0 \pm 0$  & $0 \pm 0$  & $0 \pm 0$  & $0 \pm 0$  & $0 \pm 0$  & $0 \pm 0$  \\
       & $2757 \pm 2$ & $50.3 \pm 0.2$  & (S,S,S,S,F,F)  & $1_{-0}^{+2}$  & $99_{-2}^{+0}$   & $1.0_{-0.3}^{+0.4}$  & $2.4_{-0.5}^{+0.6}$  & $50_{-1}^{+2}$  & $46.6_{-0.9}^{+0.2}$  & $0.12_{-0.03}^{+0.04}$  & $0.10_{-0.03}^{+0.04}$   & $33_{-4}^{+3}$  & $67_{-3}^{+4}$  & $0 \pm 0$  & $0 \pm 0$  & $0 \pm 0$  & $0 \pm 0$  \\
       & $2896_{-9}^{+3}$ & $57_{-28}^{+1}$  & (S,S,S,S,S,S)  & $5_{-2}^{+82}$  & $95_{-82}^{+2}$   & $3 \pm 1$  & $3_{-1}^{+11}$  & $9_{-5}^{+7}$  & $12_{-6}^{+3}$  & $40_{-8}^{+14}$  & $31_{-10}^{+1}$   & $36_{-31}^{+3}$  & $38_{-3}^{+28}$  & $12_{-2}^{+0}$  & $13_{-0}^{+5}$  & $0 \pm 0$  & $0 \pm 0$  \\
       & $2903_{-6}^{+2}$ & $34_{-12}^{+10}$  & (S,S,S,S,F,F)  & $99_{-7}^{+1}$  & $1_{-1}^{+7}$   & $10_{-6}^{+4}$  & $10_{-7}^{+6}$  & $3 \pm 2$  & $3 \pm 2$  & $29_{-0}^{+4}$  & $41_{-10}^{+22}$   & $32_{-0}^{+1}$  & $34.3_{-0.4}^{+0.1}$  & $16.8_{-0.7}^{+0.4}$  & $16.6 \pm 0.1$  & $0 \pm 0$  & $0 \pm 0$  \\[1ex]
$ 2^-$ & $2744.4 \pm 0.2$ & $236.0 \pm 0.2$  & (S,S,S,S,F,F)  & $52_{-3}^{+4}$  & $48_{-4}^{+3}$   & $6_{-1}^{+2}$  & $17.6_{-0.2}^{+0.4}$  & $22 \pm 2$  & $33 \pm 4$  & $4.4_{-0.6}^{+0.8}$  & $16.8_{-0.3}^{+0.4}$   & $26 \pm 5$  & $74 \pm 5$  & $0 \pm 0$  & $0 \pm 0$  & $0 \pm 0$  & $0 \pm 0$  \\
       & $2936_{-11}^{+3}$ & $330_{-41}^{+24}$  & (S,S,S,S,S,S)  & $100.0_{-0.2}^{+0}$  & $0.0_{-0}^{+0.2}$   & $10_{-2}^{+0}$  & $9_{-1}^{+0}$  & $15 \pm 2$  & $16_{-3}^{+1}$  & $25_{-1}^{+2}$  & $26 \pm 1$   & $9.8_{-0.4}^{+0.5}$  & $9.0_{-0.5}^{+0.8}$  & $25_{-4}^{+2}$  & $26_{-4}^{+2}$  & $15_{-2}^{+3}$  & $16_{-2}^{+3}$  \\
\end{tabular}
\end{ruledtabular}
\end{table}
\end{turnpage}

For completeness, we perform a coupled-channels calculation of the $J^P=0^-$, $1^+$ and $2^-$ $q q'\bar s \bar c$ sectors; in this case, however, the meson-meson thresholds to be included are $D^{*-}K^{+}$ (2503.94), $\bar D^{*0}K^{0}$ (2504.46), $\bar D^{0}K^{*0}$ (2760.39), $D^{-}K^{*+}$ (2761.32), $D^{*-}K^{*+}$ (2901.92) and $\bar D^{*0}K^{*0}$ (2902.40). This means that the final state $\bar DK$, through which the $T_{cs0}(2900)^0$ and $T_{cs1}(2900)^0$ resonances were found, is not reached by the tetraquark channels we are now considering. Our results are shown in Table~\ref{tab:Tcs2}. Again, many poles are found in the complex energy plane, they are all virtual states or resonances close to $\bar D^{(\ast)}K^{(\ast)}$ thresholds, with decay widths of the order of tens to hundreds of MeV. There is no evidence of bound states. The $1^+$ emerges as a promising sector for new $T_{cs1}$ states. In this case, the $\bar D^{(\ast)}K^{(\ast)}$ are in relative $S$-waves, so the formation of molecules is favoured. Potential detection channels are the lower $\bar D^*K$ channels.

\subsection{Nature of $T_{c\bar s}$ states}

In December 2022, the LHCb collaboration reported in Refs.~\cite{LHCb:2022xob, LHCb:2022bkt} a combined amplitude analysis for the decays $B^0\to \bar{D}^0 D_s^+ \pi^-$ and $B^+\to D^- D_s^+ \pi^+$, based on proton-proton collision data at centre-of-mass energies of $7$, $8$ and $13$ TeV, with an integrated luminosity of $9\,\text{fb}^{-1}$. Two new resonant states with masses and widths:
\begin{align}
T_{c\bar s0}^a(2900)^0: \hspace*{0.2cm} M &= (2892 \pm 14 \pm 15) \,\text{MeV/c}^2 \,, \nonumber \\ \Gamma &= (119 \pm 26 \pm 13) \,\text{MeV} \,, \label{eq:Tcbars0} \\[1ex] \nonumber
T_{c\bar s0}^a(2900)^{++}: \hspace*{0.2cm} M &= (2921 \pm 17 \pm 20) \,\text{MeV/c}^2 \,, \nonumber \\ \Gamma &= (137 \pm 32 \pm 17) \,\text{MeV} \label{eq:Tcbars++} \,,
\end{align}
were observed, which decay to $D_s^+ \pi^-$ and $D_s^+ \pi^+$, respectively. In Eqs.~\eqref{eq:Tcbars0} and~\eqref{eq:Tcbars++}, the first uncertainties are statistical and the second systematic. The latter state indicates the first observation of a doubly charged open-charm tetraquark state with minimal quark content $cu\bar s \bar d$, and the former appears to be its neutral isospin partner. Both states were found to have spin-parity $J^P=0^+$, and the detection channel indicates that they are isovector states, \emph{i.e.} both have isospin $1$.

Interestingly, the $cq\bar s \bar q^{\prime}$ sector was one of the first to show evidence of exotic structures, with the discovery of the $D_{s0}^*(2317)$ and the $D_{s1}(2460)$ in 2003 that did not fit the quark model expectations. These states are instead isoscalars and were tackled by our group considering them an effect of the coupling of conventional $c\bar s$ states with nearby $DK$ and $D^*K$ thresholds~\cite{Ortega:2016mms}. Since the recent $T_{c\bar s}^a$ states are isospin-1, there is no chance of coupling with naive meson structures, and a pure coupled-channels calculation has to be assumed.

Hence, we perform a coupled-channels calculation of the isospin-1 $J^P=0^+$, $1^-$ and $2^+$ $c q \bar s \bar q'$ sectors, in which the $D_s\pi$ discovery-channel can naturally be measured. We include the following meson-meson thresholds in the calculation:\footnote{In parenthesis the mass of the threshold in MeV/c$^2$.} $D_s^+\pi^-$ (2107.92), $D^0K^0$ (2362.45), $D_s^{*+}\rho^{-}$ (2887.46) and $D^{*0}K^{*0}$ (2902.40). In this case, all the quarks involved are distinguishable, but the exchange diagrams are taken into account to deal with the connection between the $DK$- and $D_s\pi$-type channels. In addition, the decay widths of the light and strange vector mesons are large enough to be taken into account. The experimental value reported in Ref.~\cite{ParticleDataGroup:2022pth} for the $\rho$-meson is $\Gamma_{\rho}=149.5\,\text{MeV}$, and the widths of the neutral and charged kaon partners are $\Gamma_{K^{\ast0}}=47.3\,\text{MeV}$ and $\Gamma_{K^{\ast\pm}}=51.4\,\text{MeV}$, respectively.

\begin{turnpage}
\squeezetable
\begin{table}[!t]
\caption{\label{tab:Tcbars1} Coupled-channels calculation of the isospin-1 $J^P=0^+$, $1^-$ and $2^+$ $c q \bar s \bar q'$ sectors ($T_{c\bar s}$ states), in which the $D_s\pi$ discovery-channel can be naturally measured. We include the following meson-meson thresholds in the calculation (in parenthesis the threshold's mass in MeV/c$^2$): $D_s^+\pi^-$ (2107.92), $D^0K^{0}$ (2362.45), $D_s^{*+}\rho^{-}$ (2887.46) and $D^{*0}K^{*0}$ (2902.40). Errors are estimated by varying the strength of the potential by $\pm10\%$.  \emph{$1^{st}$ column:} Pole's quantum numbers; \emph{$2^{nd}$ column:} Pole's mass in MeV/c$^2$; \emph{$3^{rd}$ column:} Pole's width in MeV; \emph{$4^{th}$ column:} Refers to $D_s^+\pi^-$, $D^0K^{0}$, $D_s^{*+}\rho^{-}$ and $D^{*0}K^{*0}$ Riemann sheets, respectively, with $F$ meaning first and $S$ second; \emph{$5^{th}-8^{th}$ columns:} Channel probabilities in \%; \emph{$9^{th}-12^{th}$ columns:} Branching ratios in \%.}
\begin{ruledtabular}
\begin{tabular}{lllcllllllll}
$J^P$ & $M_{\text{pole}}$ & $\Gamma_{\text{pole}}$ & RS  & ${\cal P}_{D_s^+\pi^-}$ & ${\cal P}_{D^0K^0}$ & ${\cal P}_{D_s^{*+}\rho^-}$ & ${\cal P}_{D^{*0}K^{*0}}$  & ${\cal B}_{D_s^+\pi^-}$ & ${\cal B}_{D^0K^0}$ & ${\cal B}_{D_s^{*+}\rho^-}$ & ${\cal B}_{D^{*0}K^{*0}}$\\[1ex]
\hline
$ 0^+$ & $2892_{-3}^{+4}$ & $156_{-7}^{+60}$  & (S,S,S,F)  & $0_{-0}^{+4}$  & $1_{-1}^{+9}$  & $90_{-33}^{+10}$  & $10_{-10}^{+20}$   & $43_{-38}^{+4}$  & $22_{-5}^{+2}$  & $35_{-1}^{+35}$  & $0 \pm 0$  \\
       & $2954_{-9}^{+8}$ & $129_{-11}^{+13}$  & (S,S,S,S)  & $0.5_{-0.3}^{+0}$  & $0.8_{-0.5}^{+0.3}$  & $81_{-8}^{+13}$  & $18_{-12}^{+8}$   & $6.2_{-0.8}^{+1.0}$  & $3 \pm 0$  & $54 \pm 1$  & $36 \pm 2$  \\[1ex]
$ 1^-$ & $2889_{-1}^{+2}$ & $248_{-1}^{+0}$  & (S,S,S,S)  & $0.65_{-0.07}^{+0.04}$  & $1.6_{-0.2}^{+0.1}$  & $55_{-3}^{+4}$  & $42 \pm 4$   & $25_{-5}^{+6}$  & $8_{-1}^{+2}$  & $67_{-8}^{+6}$  & $0 \pm 0$  \\[1ex]
$ 2^+$ & $2888_{-4}^{+2}$ & $155_{-5}^{+3}$  & (S,S,S,F)  & $0 \pm 0$  & $0.08_{-0.06}^{+0.08}$  & $92_{-8}^{+6}$  & $8_{-6}^{+8}$   & $0_{-0}^{+3}$  & $7_{-3}^{+90}$  & $93_{-93}^{+3}$  & $0 \pm 0$  \\
\end{tabular}
\end{ruledtabular}
\caption{\label{tab:Tcbars2} Coupled-channels calculation of the isospin-1 $J^P=0^-$, $1^+$ and $2^-$ $c q \bar s \bar q'$ sectors ($T_{c\bar s}$ states) including the following meson-meson thresholds in the calculation (in parenthesis the threshold's mass in MeV/c$^2$): $D_s^{*+}\pi^-$ (2251.77), $D^{*0}K^0$ (2504.46), $D_s^+\rho^-$ (2743.61), $D^0K^{*0}$ (2760.39), $D_s^{*+}\rho^-$ (2887.46) and $D^{*0}K^{*0}$ (2902.40). Errors are estimated by varying the strength of the potential by $\pm10\%$. \emph{$1^{st}$ column:} Pole's quantum numbers; \emph{$2^{nd}$ column:} Pole's mass in MeV/c$^2$; \emph{$3^{rd}$ column:} Pole's width in MeV; \emph{$4^{th}$ column:} Refers to $D_s^{*+}\pi^-$, $D^{*0}K^0$, $D_s^+\rho^-$, $D^0K^{*0}$, $D_s^{*+}\rho^-$ and $D^{*0}K^{*0}$ Riemann sheets, respectively, with $F$ meaning first and $S$ second; \emph{$5^{th}-10^{th}$ columns:} Channel probabilities in \%; \emph{$11^{th}-16^{th}$ columns:} Branching ratios in \%.}
\begin{ruledtabular}

\begin{tabular}{lllcllllllllllll}
$J^P$ & $M_{\text{pole}}$ & $\Gamma_{\text{pole}}$ & RS  & ${\cal P}_{D_s^{*+}\pi^-}$ & ${\cal P}_{D^{*0}K^0}$ & ${\cal P}_{D_s^+\rho^-}$ & ${\cal P}_{D^0K^{*0}}$ & ${\cal P}_{D_s^{*+}\rho^-}$ & ${\cal P}_{D^{*0}K^{*0}}$ & ${\cal B}_{D_s^{*+}\pi^-}$ & ${\cal B}_{D^{*0}K^0}$ & ${\cal B}_{D_s^+\rho^-}$ & ${\cal B}_{D^0K^{*0}}$ & ${\cal B}_{D_s^{*+}\rho^-}$ & ${\cal B}_{D^{*0}K^{*0}}$ \\[1ex]
\hline
$ 0^-$ & $2479.6 \pm 0.3$ & $134.95 \pm 0.08$  & (S,S,F,F,F,F)  & $21.99 \pm 0.04$  & $55.96_{-0.09}^{+0.08}$  & $4.14 \pm 0.01$  & $13.4 \pm 0.1$  & $1.24 \pm 0.02$  & $3.24 \pm 0.04$   & $100 \pm 0$  & $0 \pm 0$  & $0 \pm 0$  & $0 \pm 0$  & $0 \pm 0$  & $0 \pm 0$  \\
       & $2718.8 \pm 0.5$ & $371.8_{-0.3}^{+0.2}$  & (S,S,S,S,F,F)  & $0.05 \pm 0.01$  & $12.2 \pm 0.2$  & $56.06 \pm 0.10$  & $22.4 \pm 0.1$  & $0.06 \pm 0.01$  & $9.23 \pm 0.03$   & $0.9_{-0.2}^{+0.1}$  & $99.1_{-0.1}^{+0.2}$  & $0 \pm 0$  & $0 \pm 0$  & $0 \pm 0$  & $0 \pm 0$  \\
       & $2892_{-3}^{+4}$ & $251_{-4}^{+33}$  & (S,S,S,S,S,S)  & $1.1_{-0.6}^{+0.1}$  & $1.2_{-0.1}^{+0.4}$  & $20_{-12}^{+1}$  & $7_{-1}^{+0}$  & $44_{-2}^{+4}$  & $28_{-2}^{+8}$   & $23_{-3}^{+2}$  & $2.8_{-0.2}^{+0.3}$  & $33_{-4}^{+2}$  & $3_{-0}^{+3}$  & $38 \pm 7$  & $0 \pm 0$  \\[1ex]
$ 1^+$ & $2480_{-0}^{+1}$ & $59_{-3}^{+0}$  & (S,S,F,F,F,F)  & $8_{-0}^{+53}$  & $14_{-0}^{+23}$  & $2_{-2}^{+0}$  & $58_{-58}^{+0}$  & $0.4_{-0.1}^{+0}$  & $17_{-17}^{+0}$   & $100 \pm 0$  & $0 \pm 0$  & $0 \pm 0$  & $0 \pm 0$  & $0 \pm 0$  & $0 \pm 0$  \\
       & $2482.8_{-0.4}^{+0.3}$ & $26 \pm 1$  & (S,S,F,F,F,F)  & $68.5_{-0.3}^{+0.5}$  & $30.6_{-0.5}^{+0.3}$  & $0.22\pm0.01$  & $0.292_{-0.001}^{+0.002}$  & $0 \pm 0$  & $0.133 \pm 0.001$   & $100 \pm 0$  & $0 \pm 0$  & $0 \pm 0$  & $0 \pm 0$  & $0 \pm 0$  & $0 \pm 0$  \\
       & $2742_{-4}^{+2}$ & $155_{-2}^{+1}$  & (S,S,S,F,F,F)  & $0.02\pm0.01$  & $1.1_{-0.6}^{+0.8}$  & $88 \pm 7$  & $10 \pm 6$  & $0.09_{-0.04}^{+0.02}$  & $0.8_{-0.5}^{+0.6}$   & $5_{-3}^{+0}$  & $95_{-65}^{+0}$  & $0_{-0}^{+69}$  & $0 \pm 0$  & $0 \pm 0$  & $0 \pm 0$  \\
       & $2777 \pm 7$ & $115_{-7}^{+8}$  & (S,S,S,S,F,F)  & $0.51_{-0.06}^{+0}$  & $1.2 \pm 0.4$  & $74_{-6}^{+8}$  & $24_{-7}^{+6}$  & $0.32_{-0.04}^{+0.01}$  & $0.8_{-0.2}^{+0.1}$   & $5.7_{-0.2}^{+0.6}$  & $3.3_{-0.4}^{+0.5}$  & $59_{-0}^{+1}$  & $32_{-2}^{+0}$  & $0 \pm 0$  & $0 \pm 0$  \\
       & $2878_{-4}^{+3}$ & $142 \pm 1$  & (S,S,S,S,S,F)  & $0.04_{-0.01}^{+0}$  & $1_{-0}^{+1}$  & $70_{-14}^{+3}$  & $5_{-0}^{+4}$  & $21_{-4}^{+5}$  & $3_{-0}^{+2}$   & $2.3_{-0.7}^{+0.8}$  & $28_{-2}^{+3}$  & $13 \pm 2$  & $56.5_{-0.6}^{+0.1}$  & $0 \pm 0$  & $0 \pm 0$  \\
       & $2901 \pm 4$ & $83 \pm 4$  & (S,S,S,S,S,S)  & $1.2 \pm 0.1$  & $5.1_{-0.6}^{+0.3}$  & $8.6_{-0.9}^{+0.8}$  & $29_{-2}^{+4}$  & $24 \pm 4$  & $32_{-3}^{+1}$   & $18_{-3}^{+2}$  & $9.4_{-0.7}^{+0.2}$  & $13_{-2}^{+0}$  & $24_{-2}^{+0}$  & $33_{-1}^{+2}$  & $0_{-0}^{+12}$  \\[1ex]
$ 2^-$ & $2758.8_{-0.2}^{+0}$ & $338_{-2}^{+3}$  & (S,S,S,S,F,F)  & $0.19 \pm 0.01$  & $2.62_{-0.04}^{+0.03}$  & $13_{-1}^{+2}$  & $77_{-2}^{+1}$  & $0.44_{-0.06}^{+0.07}$  & $6.51_{-0.02}^{+0.01}$   & $0.8_{-0.1}^{+0.2}$  & $4.1_{-0.1}^{+0.2}$  & $95.0 \pm 0.3$  & $0 \pm 0$  & $0 \pm 0$  & $0 \pm 0$  \\
       & $2886.2 \pm 0.3$ & $296.6_{-0.7}^{+1.0}$  & (S,S,S,S,S,S)  & $0 \pm 0$  & $0.56_{-0.01}^{+0.02}$  & $1.49_{-0.07}^{+0.04}$  & $2.60_{-0}^{+0.01}$  & $47.3_{-0.5}^{+0.3}$  & $48.1_{-0.3}^{+0.4}$   & $13.1_{-0.2}^{+0.4}$  & $2.8_{-0.6}^{+0.7}$  & $47 \pm 1$  & $37 \pm 2$  & $0 \pm 0$  & $0 \pm 0$  \\
\end{tabular}
\end{ruledtabular}
\end{table}
\end{turnpage}

Table~\ref{tab:Tcbars1} shows our results. Again, with all due caution, since we should not compare Breit-Wigner resonance parameters with pole positions when we are so close to meson-meson thresholds, it is evident that we predict a resonance with quantum number $J^P=0^+$ whose mass, $2892\,\text{MeV/c}^2$, and width, $156\,\text{MeV}$, are perfectly compatible with the experimental measurements, Eqs.~\eqref{eq:Tcbars0} and~\eqref{eq:Tcbars++}. On top of that, another resonance in the same channel is found to be close to the first one; moreover, its mass and width are also compatible with the experimental measurements. It would be interesting to see if the LHCb experiment signals to one $0^+$ $T_{c\bar s}$ state or, actually, two independent resonances. There is also a singularity in each of the channels $J^P=1^-$ and $2^+$ of the scattering problem. In the first case we have a virtual state, while a resonance is found in the $J^P=2^+$ channel, though it does not seem to decay into the $D_s^+\pi^-$ final state. Both have masses close to $2.9\,\text{GeV/c}^2$ but have total decay widths larger than those of the resonances found in the $0^+$ case.

Finally, we perform a coupled-channels calculation of the isospin-1 $J^P=0^-$, $1^+$ and $2^-$ $c q \bar s \bar q'$ sectors;  in this case, however, the meson-meson thresholds to be included are $D_s^{*+}\pi^-$ (2251.77), $D^{*0}K^0$ (2504.46), $D_s^+\rho^-$ (2743.61), $D^0K^{*0}$ (2760.39), $D_s^{*+}\rho^-$ (2887.46) and $D^{*0}K^{*0}$ (2902.40). That is to say, the final state $D_s\pi$, through which the $T_{c\bar s0}(2900)^0$ and $T_{c\bar s0}(2900)^{++}$ resonances have been found, is not reached by the tetraquark channels we are now considering, but could be detected in the $D_s^{*}\pi$ or $D_s\rho$ channels. Our results are shown in Table~\ref{tab:Tcbars2}. Many poles are found in the complex energy plane. Except for one resonance state found in the $J^P=1^+$ channel with mass $2777\,\text{MeV/c}^2$ and width $115\,\text{MeV}$, all the others are virtual states mostly located near a certain $D^{(\ast)}K^{(\ast)}$ threshold, and with decay widths of the order of tens to hundreds of MeV. Again, as in the $T_{cs}$ case, the $1^+$ sector allows the $D^{(\ast)} K^{(\ast)}$ states to be in a relative $S$-wave and, thus, shows a rich spectroscopy to be worth exploring.


\section{SUMMARY}
\label{sec:summary}

The scientific community has witnessed two decades of continuous exciting discoveries of exotic hadrons through systematic searches in numerous experiments around the world. These hadrons belong mostly to the heavy quark sector and are collectively known as \emph{XYZ} states. An enormous theoretical effort has been devoted to unraveling their nature, employing a wide variety of theoretical approaches. However, due to the complexity of the problem, many of our theoretical expectations in exotic heavy hadrons are still based on phenomenological potential models.

Using data from proton-proton collisions at centre-of-mass energies of $7$, $8$, and $13$ TeV, with an integrated luminosity of $9\,\text{fb}^{-1}$, the LHCb collaboration has very recently performed amplitude analyses of the $B^+\to D^+D^-K^+$, $B^+\to D^- D_s^+ \pi^+$ and $B^0\to \bar{D}^0 D_s^+ \pi^-$ decays. For the $B^+\to D^+D^-K^+$, it is necessary to include new spin-$0$ and spin-$1$ $T_{cs}$ resonances in the $D^-K^+$ channel to obtain good agreement with the experimental data. The enhancements observed in $B^+\to D^- D_s^+ \pi^+$ and $B^0\to \bar{D}^0 D_s^+ \pi^-$ decays are interpreted as two $J^P=0^+$ $T_{c\bar s}$ states, partners of the same isospin triplet.

We have analyzed the $T_{cs}$ and $T_{c\bar s}$ states using as a theoretical framework a constituent-quark-model-based coupled-channels calculation of $qq^\prime \bar s \bar c$ and $cq\bar s\bar q^{\prime}$ tetraquark sectors. We have explored the nature and pole position of the singularities in the scattering matrix with spin-parity quantum numbers: $J^P=0^\pm$, $1^\mp$, and $2^\pm$. The constituent quark model has been widely used in heavy quark sectors and thus all model parameters are thoroughly constrained.

We find $15$ $T_{cs}$ poles in the energy range from 2.3 to 3.2 GeV/c$^2$ and a further $15$ $T_{c\bar s}$ poles in the energy interval $2.1$ to $3.0$ GeV/c$^2$. A tentative assignment of the $T_{cs0}(2900)^0$ experimental signal has been made. Our virtual state has quantum numbers $J^P=0^+$, its mass and width are $2901.9_{-0.8}^{+0.5}\,\text{MeV/c}^2$ and $51_{-1}^{+0}\,\text{MeV}$, respectively. We find a possible virtual-state candidate of the $T_{cs1}(2900)^0$ signal in the $J^P=1^-$ channel with pole parameters $2887.7_{-0.4}^{+0.3}\,\text{MeV/c}^2$ and $189.5_{-0.6}^{+0.4}\,\text{MeV}$. The theoretical width is twice the experimental value but this could be due to the fact that experimentalists simply perform cross section fits and do not derive the pole structure that produces the signals. With respect to $T_{c\bar s}$ candidates, we have predicted a resonance with quantum numbers $J^P=0^+$ and whose mass,  $2892_{-3}^{+4}\,\text{MeV/c}^2$, and width, $156_{-7}^{+60}\,\text{MeV}$, are perfectly compatible with the experimental measurements. Finally, we encourage experimentalists to search for either more $T_{cs}$ states in the $\bar D^*K$ and $\bar DK^*$ channels or $T_{c\bar s}$ signals in the $D_s^*\pi$ and $D_s\rho$ final states.


\begin{acknowledgments}
This work has been partially funded by
EU Horizon 2020 research and innovation program, STRONG-2020 project, under grant agreement no. 824093;
Ministerio Espa\~nol de Ciencia e Innovaci\'on, grant nos. PID2019-107844GB-C22 and PID2019-105439GB-C22;
and Junta de Andaluc\'ia, contract nos. P18-FR-5057 and Operativo FEDER Andaluc\'ia 2014-2020 UHU-1264517.
\end{acknowledgments}


\bibliography{PrintTcsTcsbar}

\end{document}